\documentclass{article}
\usepackage{graphicx}
\usepackage{graphicx}
\usepackage{amsmath}
\usepackage{amsfonts}
\usepackage{amssymb}
\newtheorem{theorem}{Theorem}

\newtheorem{condition}[theorem]{Condition}

\newtheorem{remark}[theorem]{Remark}

\newenvironment{proof}[1][Proof]{\textbf{#1.} }{\ \rule{0.5em}{0.5em}}

\begin{document}

\title{The paradigm of the area law and the structure of transversal and longitudinal
lightfront degrees of freedom }
\author{Bert Schroer\\present address: CBPF, Rua Dr. Xavier Sigaud 150\\22290-180 Rio de Janeiro - RJ, Brazil \\email Schroer@cbpf.br\\Prof. emer. of the FU-Berlin}
\date{May 2002}
\maketitle
\begin{abstract}
It is shown that an algebraically defined holographic projection of a QFT onto
the lightfront changes the local quantum properties in a very drastic way. The
expected ubiquitous vacuum polarization characteristic of QFT is confined to
the lightray (longitudinal) direction, whereas operators whose localization is
transversely separated are completely free of vacuum correlations. This
unexpected ''transverse return to QM'' combined with the rather universal
nature of the strongly longitudinal correlated vacuum correlations (which turn
out to be described by rather kinematical chiral theories) leads to a d-2
dimensional area structure of the d-1 dimensional lightfront theory. An
additive transcription in terms of an appropriately defined entropy related to
the vacuum restricted to the horizon is proposed and its model independent
universality aspects which permit its interpretation as a quantum candidate
for Bekenstein's area law are discussed. The transverse tensor product
foliation structure of lightfront degrees of freedom is essential for the
simplifying aspects of the algebraic lightcone holography.

\textbf{Key-words:} Quantum field theory; Mathematical physics, Quantum gravity
\end{abstract}

\section{Introduction}

Peculiarities of lightfront and ``$p\rightarrow\infty$ frame'' behavior in
particle physics have been noticed in many publications starting from the
beginnings of the 70s \cite{LKS}. In recent times we have seen a renewed
interest in the subject as a result of the idea of holography \cite{Ho} i.e.
the conjecture that for certain geometric constellations it may be possible to
encode degrees of freedoms and most of the properties of a QFT in d-spacetime
dimensions into a suitably chosen lower dimensional geometric carrier.
Although the first intuitive picture about such encoding came from consistency
observations on rotational symmetric black holes, there were also arguments
that a similar holographic encoding may occur in plane Minkowski space
lightfront physics \cite{Suss}; in fact such an idea receives additional
support from the analogy of the Hawking effect caused by black holes with
bifurcated horizons with the Unruh effect associated with the Rindler wedge in
Minkowski spacetime \cite{Wald} in the sense that if there exists an analogy
on the thermal side of a Hawking temperature and that seen by an Unruh
observer, there should also be an analogous behavior of other thermodynamic
aspects as e.g. a suitably defined ''localization entropy'' associated to the
horizon (i.e. half of the lightfront). Since a thermal interpretation in terms
of fundamental laws imposed on the black hole situation requires the
Bekenstein area law for entropy, and since the finite surface of a black hole
corresponds to the infinite surface defined by the edge of the Rindler wedge,
one would expect to find a rather universal constant surface density of the
entropy of the vacuum which in its restriction to a suitably defined
subalgebras on the (upper) horizon of the Rindler wedge is described by a
density matrices.

The main aim of this note is to present new concepts and mathematical tools
which show that this is not only an analogy between special situations in
curved spacetime and some new aspects of lightfront physics (as compared to
the way the lightfront and the $p\rightarrow\infty$ frame method were
previously used in particle physics), but rather the start of a paradigmatic
change in looking at local quantum physics\footnote{There are indications for
an ongoing paradigmatic change in various other areas of QFT \cite{Re}%
\cite{Bu}\cite{Hol}\cite{Br}.}. In order to appreciate this statement the
reader is reminded that for several decades there exist two ways of dealing
with QFT which are mainly different in their interpretation, mathematical
implementation of concepts and underlying philosophy but which were based on a
shared stock of principles. The standard approach (i.e. that of most textbooks
and used by the vast majority of researchers) attributes a direct physical
reality to pointlike quantum (suitably averaged in the sense of Bohr and
Rosenfeld) fields, whereas the operator-algebraic setting is based on the idea
that joint properties of algebras generated by operators which share the same
spacetime localization already contain all the physics (analogous to the
fundamental role of localization of events in with particle counters whose
inner working is in most cases unknown \cite{Haag}), but in order to unravel
it one cannot use the standard methods.

The historically first observation which lend some respectability to the
algebraic viewpoint is the natural explanation of the insensitivity of the
S-matrix against local changes of the field-coordinatization. The support from
the S-matrix viewpoint was strengthened by the recent observation that unitary
crossing symmetric (this being the on-shell substitute for the missing
off-shell Einstein causality) have maximally one system of local algebras if
they have one at all \cite{S-matrix}. However in many cases these observations
which deemphasize the role of individual fields in favour of local equivalence
classes really do not require the use of the algebraic viewpoint since e.g.
the S-matrix continues to be expressible by LSZ formulas in terms of pointlike
fields. This is quite different in the present case of lightfront holography
\cite{Hor}\cite{S2}.

Although we are aware that the general public level of knowledge about modular
theory, on which this paper is based, is lagging far behind its importance in
local quantum physics, this short note is not the place for looking at
background information. For attempts in this direction and comments on its
interesting history we rather refer to relevant sections in \cite{Haag} and
\cite{Bo-JMP}.

The present algebraic lightfront holography incorporates concepts which have
attracted attention in various areas of particle physics, curved spacetime QFT
\cite{Wald}, basic quantum physics and quantum measurement \cite{qu} and even
in physics-based philosophical contributions \cite{Hal}.

For the convenience of the reader we have collected the previously presented
results on lightfront restrictions of free fields as well as some geometric
definitions concerning lightfront, wedges and their lightfront-horizon
\cite{S2}\cite{Hor} in compressed form in an appendix. The reader is strongly
advised to have a look at the results obtained by this mathematically more
elementary lightfront restriction method (which is limited to free fields)
before he reads the next section.

\section{The setting of local quantum physics and its adaptation to the lightfront}

In the derivation of the time-dependent scattering theory and the analytic
properties carried out in the 50s and 60s it became clear that pointlike
fields are analogous to coordinates in differential geometry. These
investigations also increased the awareness about the limitations of the
adaptation of the classical field picture to local quantum physics
(''quantization'') and, in particular, suggested that the attribution of a
physical reality to individual fields may be a bit of an illusion which in
certain cases may even conceal basic intrinsic properties. This is indeed the
case for QFT on the lightfront as we hope to make clear in the sequel.

It is assumed that the reader knows some basic facts about the algebraic
approach which describes QFT in terms of a collection (net) of spacetime
indexed operator algebras \cite{Haag} in a common Hilbert space fulfilling
causality- covariance- and spectral- properties. These properties are
adaptations of those which already appeared in the first non-Lagrangian
setting of QFT by Wightman \cite{St-Wi} and which nowadays are often referred
to as the linear Wightman requirements. Besides the linear Einstein causality
(spacelike local commutativity) there are also causality requirements which
are an algebraic substitute for an equation of motion namely the causal shadow
property or primitive causality \cite{H-S}
\begin{equation}
\mathcal{A}(\mathcal{O})=\mathcal{A}(\mathcal{O}^{\prime\prime})
\end{equation}
i.e. the operator algebra (always weakly closed) localized in a simply
connected region $\mathcal{O}$ equals that of its causal completion
($\mathcal{O}^{\prime\prime}$ is the spacelike disjoint of $\mathcal{O}%
^{\prime}$ which in turn is the spacelike disjoint of $\mathcal{O}$). Such
properties have no natural expressions in terms of the mainly linear Wightman
properties and rather require the setting of operator algebras \cite{H-S}.
Since we will also be concerned with subalgebras associated with
lower-dimensional regions as, e.g., lightfront algebra $\mathcal{A}(LF)$ we
will also assume the \textit{characteristic extension of causal shadow
properties} of which the following relation is the most important case
\begin{equation}
\mathcal{A}(W)=\mathcal{A}(LF_{+})\label{char}%
\end{equation}
Classically its content and validity should be obvious: the field data in the
(Rindler) wedge\footnote{Our reference wedge will be always the $x^{0}-x^{1}%
$-wedge, the wedge in any other position is obtained by applying Poincar\'{e}
transformations.
\par
{}} $W=\left\{  x|\,\left|  x^{0}\right|  <x^{1},x^{1}>0\right\}  $ are
determined in terms of the characteristic data on its upper causal horizon
$LF_{+}$ (the $x_{{}}^{1}>0$ half of the light front). $W$ is the causal
shadow of $LF_{+}$ since every particle or lightray which has passed through
$LF_{+}$ must have before passed through $W$ and vice versa$.$ But note that a
region on $LF$ which is bounded in the $x^{1}$ direction does not cast any
causal shadow at all; this is one of the peculiarities of the lightfront. In
QFT it has been shown for free fields \cite{Driessler}\cite{GLRV} and holds
(within the present level of mathematical rigor\footnote{The present
constructive methods for d=1+1 factorizable models deal with bilinear forms
(matrix elements of ``would be'' operators) and fall still short of genuine
operator algebra methods.}) in the algebraic formulation of d=1+1 factorizing
models \cite{Schroer}. Since it is natural extension of the causal shadow
principle (limiting case of the spacelike causal shadow principle of local
quantum physics \cite{Haag}) and fits perfectly into the modular setting, we
add it to the list of requirements of local quantum physics leaving open the
very plausible possibility that it may turn out to follow already from the
commonly accepted principles. The only exception for the validity of
(\ref{char}) are d=1+1 chiral theories.

The operator algebra on the left hand side in (\ref{char}) is defined in terms
of the net-setting of AQFT as all operator algebras associated with noncompact
localization regions namely
\begin{equation}
\mathcal{A}(W)=\overline{\bigcup_{\mathcal{O}\subset W}A(\mathcal{O})}%
\end{equation}
where the closure is in the weak operator topology. Such observable algebras
are required to obey the geometric Bisognano-Wichmann property $\mathcal{A}%
(W^{\prime})=\mathcal{A}(W)^{\prime}$ where $W^{\prime}$ is the geometric
opposite (spacelike disjoint) of $W$ and the dash on the operator algebra
denotes as usual the von Neumann commutant algebra \cite{Haag}. Wedge algebras
are actually von Neumann factors i.e. $\mathcal{A}(W)\cap\mathcal{A}%
(W)^{\prime}=\mathbb{C}\mathbf{1,}$ $\mathcal{A}(W)\vee\mathcal{A}(W)^{\prime
}=B(H)$ \cite{Bo-JMP}. The upper horizon algebra $\mathcal{A}\left(
LF_{+}\right)  $ on the right hand side is best defined in terms of the result
of the following theorem on modular inclusion:

\begin{theorem}
(Wiesbrock, Borchers \cite{Wies}\cite{Bo-JMP}) Let W$_{e_{+}}\subset\,$W be
the lightlike translated wedge algebra e$_{+}$=(1,1,0,0)$.$ The inclusion of
operator algebras
\begin{equation}
\mathcal{A}(W_{e_{+}})\equiv AdU(e_{+})\mathcal{A}(W)\subset\mathcal{A}(W)
\end{equation}
is ``modular'' i.e. the modular unitary $\Delta_{W}^{it}$ of the standard pair
($\mathcal{A}(W_{e_{+}}),\Omega$) ``compresses'' the smaller algebra
\begin{equation}
\sigma_{t,W}(\mathcal{A}(W_{e_{+}}))\equiv Ad\Delta_{W}^{it}\mathcal{A}%
(W_{e_{+}})\subset\mathcal{A}(W_{e_{+}}),\,t>0
\end{equation}
In this case the original positive energy lightray translation U$_{e_{+}}%
$(a)$\equiv$U(ae$_{+}$) can be recovered from the two modular unitary groups
$\Delta_{W}^{it}$,$\,\Delta_{W_{e_{+}}}^{it},$ and the lightlike translation
$U_{e_{+}}(a)$ together with the modular group $\Delta_{W}^{it}.$ obey the
Borchers translation-dilation commutation relation
\begin{equation}
\Delta_{W}^{it}U_{e_{+}}(a)=U_{e_{+}}(e^{-2\pi t}a)\Delta_{W}^{it} \label{Bo}%
\end{equation}
If the relative commutant
\begin{equation}
\mathcal{A}(W_{e_{+}})^{\prime}\cap\mathcal{A}(W)
\end{equation}
is also standard with respect to the vacuum $\Omega$ (in which case the
modular inclusion is called ``standard''\footnote{The terminology ``standard''
refers to a pair (algebra, reference state vector)=($\mathcal{A},\Omega$) and
stands for the cyclicity ($\overline{\mathcal{A\Omega}}=H$) and the separating
property ($A\Omega=0,A\in\mathcal{A}\curvearrowright A=0$). In QFT the
reference state is usually the vacuum vector and the standardness is called
the Reeh-Schlieder property (or somewhat imprecisely ``the
operator-statevector relation'').})$,$ then the two definitions
\begin{align}
\mathcal{A}(LF(0,1))  &  \equiv\mathcal{A}(W_{e_{+}})^{\prime}\cap
\mathcal{A}(W)\label{horizon}\\
\mathcal{A}(LF_{+})  &  \equiv\overline{\bigcup_{a>0}AdU_{e_{+}}%
(a)\mathcal{A}(LF(0,1))}\nonumber\\
\mathcal{A}(LF_{-})  &  =J\mathcal{A}(LF_{+})J\nonumber
\end{align}
lead to%
\begin{align*}
\mathcal{A}(LF_{+})  &  =\mathcal{A}(W)\\
\mathcal{A}(LF_{-})  &  =\mathcal{A}(W)^{\prime}=\mathcal{A}(W^{\prime})
\end{align*}
Here $J$ is the modular involution of the standard pair ($\mathcal{A}%
(W),\Omega).$ The resulting net structure on the lightfront algebra
$\mathcal{A}(LF)$ is that of a generalized chiral theory; in addition to
invariance under lightray translation and dilations (\ref{Bo}) the vacuum is
invariant under a positive generator $L_{0}$ rotation \cite{Guido} whose
action on the original net is ``fuzzy'' \cite{S1} (not representable by a diffeomorphism).
\end{theorem}

This theorem requires several comments. In its present form it is the
adaptation of an abstract mathematical theorem on modular inclusion to local
quantum physics. In fact it comprises three closely intertwined theorems, the
related Borchers- and Wiesbrock- theorems \cite{Bo-JMP} and a theorem on the
equivalence of standard modular inclusions with generalized chiral QFTs
\cite{Guido}.

The standard modular inclusion theorem resolves the problem of the
longitudinal localization structure of the lightfront algebra, but it is
``blind'' with respect to the transverse direction of the edge of the
bifurcated horizon of a Rindler wedge. It is precisely the presence of these
transverse degrees of freedom\footnote{The ``degree of freedom'' terminology
and their counting in the calculation entropy makes literal sense only for
free (or systems with quanta similar to free systems) quantum field systems;
its meaning outside is of a more physical intuitive nature, which receives
some support from the phase space structure of local quantum physics
\cite{Haag}.} which distinguish the generalized chiral theories of the
previous theorem from those more familiar standard chiral theories which
result from the chiral tensor decomposition of d=1+1 conformal theories. The
former permit the nontrivial action (transverse translations and rotation) of
automorphisms which, in lack of a transverse localization structure, are at
first sight hard to distinguish from internal symmetries. However, in the next
section, we will show how to construct a transverse localization with quite
unexpected properties (alluded to in the abstract) using the methods of AQFT.
The relevant Poincar\'{e} transformations are those which appeared before in
the attempts to reconstruct the original massive net from its holographic
lightcone projection .

Let us collect those properties in form of a condition on which our analysis
relies but which were not part \ of Wightman's framework \cite{St-Wi} of what
constitutes a QFT.

\begin{condition}
(extended causal shadow property and cyclicity condition) Physically
admissable models of local quantum physics fulfill in addition to the standard
linear properties (local commutativity, covariance and spectral positivity)
also the causal shadow property\footnote{This condition does not contradict
the well-known limitation of the canonical equal time formalism since the
double-cone algebras (which become associated with an equal time base via this
causal shadow property) admit generally no generating equal time fields. The
situation on the lightfront is totally different.} including its extension
(\ref{char}) as well as the cyclicity (Reeh-Schlieder property) of all
relative commutants (even if they the localization region, as it happens e.g.
for compact regions on the lightfront, do not cast a causal shadow i.e. are
identical to their causal completion) as used in the previous modular
inclusion theorem.
\end{condition}

These causality properties allow to geometrically identify (even arbitrarily
tight-localized) subalgebras which fulfill the Reeh-Schlieder cyclicity; they
turn out to be essential for the formulation of lightfront holography. Whereas
in quantum mechanics (say in the multiplicative form of second quantization in
order to facilitate the comparison with QFT) a quantization box splits the
world into a tensor product between the inside and outside quantum mechanical
Fock spaces in such a way that the vacuum factorizes without entanglement, the
QFT situation is the extreme opposite: from the algebra of an arbitrary small
spacetime region (e.g. a small double cone as the prototype of a relativistic
box) one can approximate any state of the world from the vacuum. The denseness
of the subspace cyclically generated from the vacuum is always guarantied as
long as the open localization region has a nontrivial causal disjoint.
Accepting this condition as a characterization of ``physically admissable'' is
a reasonable working hypotheses (all the QFT which have been used fall into
this category) and should not be confused with installing a doctrine.
Stringlike QFTs as Wigner's ``helicity towers'' (QFTs associated with zero
mass faithful little group representations which contain all helicities
including the helicity of the graviton $h=2$) which do not permit a nontrivial
localization in compact regions are of course legitimate objects of further
research. This is especially so since their quantum consisteny requires the
semiinfinite stringlike localization in any spacetime dimensions (more
precisely the spacelike cone localization of algebraic QFT), but does not
bring about restrictions to d=10, 24 spacetime dimensions. The reason for the
disharmony between the causality and stability principles of QFT and string
theory is not that string theory invents structures which are new and untested
but rather that its premature and unfounded claim of being a theory of
everything precludes the necessary work to confront the principles of QFT and
in this way reveal (as it was done with all physically relevant new theories
before) its possible new aspects.

A very important mathematical tool in the ongoing paradigmatic change of QFT
\ is the Tomita-Takesaki modular theory of operator algebras. \textit{All the
significant differences to QM may be traced back to the phenomenon of vacuum
polarization} which in turn is a direct consequence of the causality and
spectral stability properties (positivity of timelike translation generators).
The omnipresence of vacuum polarizations and the related spacelike
correlations in vacuum expectations of products of spacelike separated
observables even in interaction-free theories is a well-known phenomenon ever
since Heisenberg discovered it through studying the global limit of local
Noether charges in free field theories in 1934 \cite{Hei}.

A closely related refinement of the manifestation of vacuum polarization which
is able to detect the presence of interactions (and therefore is of direct
relevance to particle physics) is expressed in the following recent theorem on
existence of properties of ``polarization-free generators'' (PFG) of localized
one-particle creation operators. A PFG $G$ is an (generally unbounded)
operator affiliated with an operator algebra $\mathcal{A}(\mathcal{O})$ (in
mathematical notation) $G\eta\mathcal{A}(\mathcal{O})$ such that (\# denotes
as usual either no superscript or the hermitian conjugation)
\[
G^{\#}\Omega=1-particle\,\,vector
\]

\begin{theorem}
(\cite{BBS}) Wedge-localized operator algebras in theories describing massive
particles always possess affiliated PFGs F%
\[
G\eta\mathcal{A}(W)
\]
whereas the existence of PFGs for algebras localized in causally closed
subwedge regions imply the absence of interactions in the subspaces (sectors)
generated by repeated application of $G$ to the vacuum$;$ in fact in this case
the translates of PFGs $G(x)=AdU(x)G$ are actually linearly related to free fields.
\end{theorem}

This theorem is a strengthened form within the operator algebra setting of an
old well-known theorem which characterizes free fields in terms of two-point
functions \cite{St-Wi} and which recently among other things turned out to be
of interest in connection with the protection mechanism in conformal SYM
theories \cite{JPL}.

These aspects of PFGs are interesting for two reasons. On the one hand one
obtains an entirely intrinsic characterization of absence/presence of
interactions which does not depend on the chosen field coordinatization in the
equivalence class of all local fields\footnote{Since the Wick-polynomials of
free fields have quite complicated correlation functions, it would be
difficult to assert the absence of interactions by just looking at them.}. On
the other hand the special role attributed to wedge algebras may be used as
the start of a new constructive approach to QFT. Although the interaction is
at first sight not directly visible on the level of the wedge algebra
generated by wedge-affiliated PFGs, it makes its appearance in the properties
of the modular operators \cite{Schroer}. The TCP-related modular involution
$J$ turns out to differ from its interaction-free form by the appearance of
the scattering operator $S_{scat},$ which thereby acquires a new role related
to localization (which completely escaped the old S-matrix philosophy
according to which on-shell quantities do not reveal localization properties).
Although it has up to now not been possible to show the existence of a QFT
associated with a given admissable (unitary, crossing symmetric with the
necessary analytic properties) S-matrix, the application of modular theory
does lead to the uniqueness of the associated would be AQFT, i.e. there is
either none or just one local quantum theory with an admissable S-matrix
\cite{S-matrix}. Knowing the structure of the wedge algebra one may construct
the operator algebras of smaller regions by forming algebraic intersections
(instead of restricting the support of smearing functions as one would
restrict localization regions in the standard approach). That this completely
intrinsic algebraic way is practicable has been shown recently in some ``well
behaved'' but nevertheless nontrivial two-dimensional models of factorizable
models. The mysterious nonlocal Zamolodchikov-Faddeev algebra which appeared
as a computational tool acquires for the first time a spacetime
interpretation: its Fourier transforms are the PFGs of the wedge algebra
\cite{Schroer}.

\section{Tranverse Localization on the Lightfront and the return of
vacuum-polarization-free Quantum Mechanics}

We now turn to the issue of equipping the transverse lightfront directions
with a localization structure\footnote{In the case of free-field generated
algebras the structure of the lightfront algebras can be simply obtained by
restricting the underlying free fields to the lightfront \cite{Driessler}, and
the use of the modular inclusion techniques and the lightlike Wigner little
group would be unnecessary.}. The modular inclusion techniques of the previous
section has provided us with a chiral net. But it would be wrong to think of
these horizon-affiliated chiral nets as being of the same kind as those
families which one has found by chiral tensor decomposition of d=1+1
dimensional conformal theories. Even though modular theory suggests strongly
that any Moebius-covariant chiral net contains a representation of the
diffeomorphism group of the circle (but without the possibility of defining a
physical energy-momentum tensor through the Witt-Virasoro infinitesimal
generators $L_{n}$), there is absolutely no reason to expect the existence of
a finite trace (character) of the Gibbs operator $e^{-\beta L_{0}}$ since the
transverse translational symmetry would force each potential eigenvalue to be
infinitely degenerate. With other words one is dealing with a generalized
chiral situation in which one cannot expect the important ``split property''
\cite{Haag} to hold\footnote{In fact one can show directly that the transverse
symmetry wrecks the split property in the lightray direction, i.e the fuzzy
localized intermediate type I factors cannot exist.} even if it was valid in
the original d-dimensional ambient theory. This transverse alignment of
degrees of freedom is just another illustration of the rather radical aspects
of algebraic holography which are the prize to be paid for the gain in
kinematical simplicity and universality of certain algebraic structures. It is
the main reason why measures of the relative cardinality of quantum field
theoretical situations in which the thermal aspects are caused by localization
exhibit an an area law instead of the volume proportionality of entropy
associated with standard heat bath thermal physics.

Hence the modular inclusion construction leads to a \textit{generalized}
chiral theory whose longitudinal interval-associated net $\left\{
\mathcal{A}(I)\right\}  _{I\subset\text{{R}}}$ (where {\text{\r{R}}} denotes
the compactified lightray) should be pictured as the operator algebra of a d-2
dimensional strip of width $I$ in longitudinal direction$.$ We want to create
a transversal net structure i.e. construct algebras $\mathcal{A}%
(\mathcal{S(O}_{\bot}\mathcal{)})$ belonging to the two-sided longitudinal
extension of a d-2 dimensional compact region on the edge of the horizon (or
of the wedge $W$). If we would have a net of algebras associated with compact
regions $\mathcal{O}_{LF}\subset LF$ on the d-1 dimensional lightfront we
could form these strip algebras by the weak closure of algebras $A(\mathcal{O}%
_{LF})$ in the usual manner
\begin{equation}
\mathcal{A}(\mathcal{S(O}_{\bot}\mathcal{)})=\overline{\bigcup_{O_{LF}%
\subset\mathcal{S(O}_{\bot}\mathcal{)}}\mathcal{A}(\mathcal{O}_{LF})}
\label{long}%
\end{equation}
Such a net structure can indeed be obtained via applying Wigner's little group
transformations which preserves the direction of the horizontal lightray. In
fact the idea was first used in a joint paper with Wiesbrock \cite{S-W}
(chiral ``scanning'') in order to enrich the standard modular inclusions by
forming intersections (``modular intersections'') of the original $W$ with
L-tilted wedges $W_{i}$ which share the upper defining lightray and as a
consequence also the upper horizon. Projected into the lightfront the
translation part of the Wigner little group is represented by a
Galilei-transformation $G_{v}$%

\begin{equation}
\mathbf{x}_{\perp}\rightarrow\mathbf{x}_{\perp}\mathbf{+v}x_{+},\;\,x_{+}%
\rightarrow x_{+}%
\end{equation}
The d-2 dimensional strip algebra $\mathcal{A}(I)$ in the direction of the
transverse bifurcation edge is transformed into an inclined position within
the lightfront. The compact intersections of the original with the transformed
strip leads to local algebras (which according to the condition of the
previous section are nontrivial algebras)%
\begin{equation}
\mathcal{A}(\mathcal{S(O}_{\bot}\mathcal{)})\cap\mathcal{A}(G_{v}%
(\mathcal{S(O}_{\bot}\mathcal{))})
\end{equation}
which are the desired building blocks for the compact $\mathcal{O}_{LF}$
localization region used for defining the net of operator algebras
$\mathcal{A}(\mathcal{O}_{LF})$ on the lightfront from which in turn the
longitudinal strip algebras (\ref{long}) may be constructed. Following
Driessler we now show that the foliation of the lightfront algebra
$\mathcal{A}(LF)$ into nonoverlapping commuting longitudinal strip algebras%

\begin{align}
\mathcal{A}  &  =\mathcal{A}(LF)=\overline{\bigcup_{i}\mathcal{A}%
(\mathcal{S}(\mathcal{O}_{\bot}^{(i)}))}\\
LF  &  =\bigcup_{i}\mathcal{S}(\mathcal{O}_{\bot}^{(i)}),\,\,\mathcal{O}%
_{\bot}^{(i)}\cap\mathcal{O}_{\bot}^{(j)}=\emptyset,\,i\neq j\nonumber
\end{align}
has a tensor product structure i.e. fulfills the strongest form of statistical
independence which quantum physics is able to offer.

\begin{theorem}
The transverse foliation of the lightfront algebra is a tensor-product
factorization
\begin{align}
\mathcal{A}(LF)  &  =\bar{\otimes}_{i}\mathcal{A}(\mathcal{S(O}_{\perp}%
^{(i)}\mathcal{)})\label{fac}\\
H  &  =\bar{\otimes}_{i}H_{i},\,\,\Omega=\bar{\otimes}_{i}\Omega
_{i}\nonumber\\
H_{i}  &  =\overline{\mathcal{A}(\mathcal{S(O}_{\perp}^{(i)}\mathcal{))}%
)\Omega}\nonumber
\end{align}
\end{theorem}

\begin{proof}
The proof is based on the existence of a transverse ``get away'' shift, i.e.
by translating the strip transversely outside itself we obtain according to
spacelike local commutativity two commuting strip algebras and the rather
surprising statement that there exists no transverse vacuum polarization.
According to an old result of \cite{Borchers} one knows that each such algebra
has a (lightlike) translation-invariant center and the lightlike translation
acts as an inner automorphism. We use the projector onto the subspace which
the strip algebra generates cyclically from the vacuum. Now look at the
analytic properties associated with the following relation between an operator
$A\in\mathcal{A}(\mathcal{S(\mathcal{O}_{\bot}}))$ and another one from the
commutant $A^{\prime}\in\mathcal{A}(\mathcal{S(\mathcal{O}_{\bot}}))^{\prime}$%
\begin{equation}
\left\langle 0\left|  AU_{e_{+}}(a)A^{\prime}\right|  0\right\rangle
=\left\langle 0\left|  A^{\prime}U_{e_{+}}^{\ast}(a)A\right|  0\right\rangle
\end{equation}
But using the analyticity due to the positivity of the generator, we obtain
via the Liouville theorem the constancy in $a$ and hence
\begin{equation}
\left\langle 0\left|  AA^{\prime}\right|  0\right\rangle =\left\langle
0\left|  A\right|  0\right\rangle \left\langle 0\left|  A^{\prime}\right|
0\right\rangle
\end{equation}
which together with the cyclicity condition of the previous section leads to a
type $I_{\infty}$ tensor factorization, as well as the stronger statement that
the vacuum has no entanglement with respect to the inside-outside tensor
factorization. Recursive application together with the fact that
(\ref{horizon}) $\mathcal{A}(LF)=\mathcal{A=B(}H\mathcal{)}\,$\ yields the
tensor factorization and absence of vacuum entanglement in the tiling of the
lightfront by longitudinally directed strips i.e. the equations (\ref{fac}).
The one-sided strip algebras $\mathcal{A}(\mathcal{S}_{+}\mathcal{(\mathcal{O}%
_{\bot}}))$ on the horizon $LF_{+}$ are easily shown to be type hyperfinite
type $III_{1}$ algebras within the respective tensor factors so that
$\mathcal{A}(\mathcal{S(\mathcal{O}_{\bot}}))=\mathcal{A}(\mathcal{S}%
_{-}\mathcal{(\mathcal{O}_{\bot}}))\vee\mathcal{A}(\mathcal{S}_{+}%
\mathcal{(\mathcal{O}_{\bot}}))$. Since the argument is the same as that for
wedges, it will not be repeated here \cite{Bo-JMP}.
\end{proof}

This state of affairs requires some comments. In local quantum physics one
encounters several situations in which the notions of ``causal disjoint''
becomes synonymous with ``nonoverlapping''. For standard chiral theories on
$S^{1}$ this is well-known, but it also holds for conformal observables living
in the Dirac-Weyl compactified Minkowski spacetime if ``nonoverlapping'' is
interpreted as ``nonoverlapping of the lightlike prolongation'' (Huygens
principle) \cite{G-L}\cite{S}. Via the CQFT-AdS isomorphism \cite{Re} this
simplified causality picture is inherited by observables in anti deSitter
spacetime. But none of these cases has such far-reaching factorization
properties as the local quantum physics on the lightfront and (see next
section) lightcones. This is why the metaphoric sounding terminology `` return
of QM'' was used in the title of this section.

The usefulness of auxiliary constructs and associated (holographic)
reprocessing of degrees of freedom depends very much on how the degrees of
freedom ``align'' themselves after they received their new
``spacetime-indexing''. In this respect the holographic reprocessing onto
causal horizons (for the rotational case see next section) is extraordinarily
rich because it preempts the transversal factorization structure which is
necessary for area proportionality of vacuum-split caused entanglement entropy
(i.e. entropy per unit transversal volume).

In order to convert this multiplicative transverse composition law of
lightfront degrees of freedom into an additive area law we should search for
an appropriate notion of entropy in which the vacuum turns into a temperature
state. This certainly does not happen relative to the full quantum mechanical
strip algebras $\mathcal{A}(\mathcal{S(\mathcal{O}_{\bot}}))$ relative to
which the vacuum remains a pure state. The situation changes if by cutting the
strip into two pieces we restrict the vacuum to the horizon component
$\mathcal{A}(\mathcal{S}_{+}\mathcal{(\mathcal{O}_{\bot}})).$ In this case the
vacuum state turns into a temperature state at the Hawking temperature
$T=2\pi$ (in appropriate units) but since this algebra is easily shown to be a
hyperfinite type III$_{1}$ factor (for the same reason as the wedge algebras
\cite{Bo-JMP}) this is bad for the definition of an entropy since such
algebras simply do not permit to associate an entropy to any state. The reason
for this are the uncontrollable vacuum fluctuations at the boundary, a
phenomenon well-known from the divergencies in the definition of partial
Noether charges involving a sharp spatial cutoff as mentioned in the previous
section. As in that case the choice of smearing functions which have the
constant value one in the desired localization region of the partial charge
and zero outside a slightly bigger region (thus leaving a ``collar'' inside
which the fluctuation can be controlled) the split property achieves something
similar for algebras. Given two hyperfinite type III$_{1}$ algebras, one
sharply localized in a causally closed double cone $\mathcal{O}$ and the other
in $\mathcal{O}+\delta$ (where $\delta$ is the collar size). If there exists
an intermediate quantum mechanical algebra $\mathcal{N}$ the inclusion is
called the ``split'' \cite{Haag}. In this case one achieves a control of
vacuum polarizations through the use of the quantum mechanical type I factor
$\mathcal{N}$ between $\mathcal{A}(\mathcal{O})$ and $\mathcal{A}%
(\mathcal{O}+\delta)$%
\begin{align}
\mathcal{A}(\mathcal{O})  &  \subset\mathcal{N\subset A}(\mathcal{O}+\delta)\\
\mathcal{A}  &  =B(H)=\mathcal{N}\bar{\otimes}\mathcal{N}^{\prime}\nonumber\\
H  &  =H_{in}\bar{\otimes}H_{out}\nonumber
\end{align}
The last two lines express the tensor factorization of the algebra into
``inside and outside degrees of freedom i.e. those contained in $\mathcal{A}%
(\mathcal{O})$ and those in $\mathcal{A}(\mathcal{O}+\delta)^{\prime}$ i.e.
localized ``outside'' (meaning in the relativistic setting: causally disjoint
from) $\mathcal{O}+\delta.$ This is the best analogy of the ``inside/outside''
tensor factorization of (second quantized) quantum mechanics which the
presence of vacuum polarization and the causal shadow property permits. It has
turned out that contrary to the aforementioned Noether charge analogy which
depends on the form of the smearing function in the collar region, the
intermediate fuzzy-localized (between $\mathcal{O}$ and $\mathcal{O}+\delta$)
algebra $\mathcal{N}$ is canonically (more precisely functorially, \cite{D-L})
related to the given pair. The restriction of the vacuum to $\mathcal{N}$ is
not only a thermal state at a Hawking temperature, but it also permits a
conventional density matrix description%
\begin{align}
\omega|_{res}  &  \rightarrow\rho,\,\,\rho>0,\,\,tr\rho=1\\
S_{ent}  &  =-tr\rho\ln\rho\nonumber
\end{align}
which is the prerequisite for defining a von Neumann entropy. Clearly such an
entropy would have a geometric localization aspect (i.e. it would represent an
entropy associated with the localization-caused temperature) as well as a
quantum interpretation in form of a reduction of the vacuum entanglement
($\Psi_{i},\Phi_{j}$ basisvectors in $H_{inn},H_{out}$) to the subsystem
$\mathcal{N}$.
\begin{align}
\Omega &  =\sum_{i,j}a_{ij}\Psi_{i}\bar{\otimes}\Phi_{j}\label{split}\\
\rho &  =(\rho_{ik})\text{ }in\,\text{\thinspace}\Psi-basis\,\,of\,\,H_{inn}%
\nonumber\\
\rho_{ik}  &  =\sum_{j}a_{ij}a_{kj}^{\ast}\nonumber
\end{align}
In order to have a finite entropy the density matrix $\rho$ should not have
too many eigenvalues near zero.

The clarity of these concepts and their mathematical formulations stands in
contrast to the difficulties one encounters in attempts to calculate or
estimate $\rho$ and the associated entropy along these lines$.$ Even in the
simplest setting of chiral theories this has not been achieved. There is
another definition of entropy which uses the same physical intuition but whose
mathematical relation to (\ref{split}) is not known. This is Araki's relative
entropy between two states on the same algebra in our case between the vacuum
state $\omega$ and the product vacuum $\omega_{p}$ \cite{Narn} on
$\mathcal{A}(\mathcal{O})\vee\mathcal{A}(\mathcal{O}+\delta)^{\prime}$. Kosaki
found a beautiful variational representation of this relative entropy which
avoids the use of modular theory and uses only the two states. Again this
entropy unfortunately resisted up to now calculational attempts for the
problem at hand \cite{Narn}. There are other calculational more accessible
formulas for localization entropy which are less canonical and (analog to the
testfunction dependence of the above partial charge) dependent on how one
realizes the two states. Intuitively one expects that the leading behavior for
$\delta\rightarrow0$ is the same, but a rigorous proof for this is still
missing. One such family of formulas uses the connection between tensor
product doubling and the introduction of doublet fields \cite{ADFL}. In this
doublet description the unitary operators $U(f)$ which implements the split
isomorphism and transforms the vacuum into its split version can be written in
terms of a Noether current $j$ defined by the doublet integrated with
testfunctions $U(f)=e^{ij(f)}$ subject to certain support requirements (in
analogy to the partial charge, the $f$ is chosen such that in $\mathcal{O}$
one has $f=1,$ whereas in the causal disjoint $f$ vanishes). For the split
algebra generated by a free chiral Fermion this suggests a linear vanishing
with decreasing collar size $\delta\rightarrow0$ \cite{S2}$.$ Since the split
situation in this limit is known to become inequivalent to the vacuum
representation of the limiting theory $\mathcal{A}(\mathcal{O})\vee
\mathcal{A}(\mathcal{O}^{\prime}),$ this vanishing property is inherited by
all vectors which are cyclically generated by applying operators from the
algebra to the doubled vacuum $\Omega\bar{\otimes}\Omega$ and its $U(f)$
transform. With all overlaps approaching zero linearly with vanishing distance
$\delta,$ one naturally expects the linear vanishing of the $a_{ik}$
coefficients in the decomposition of the vacuum in the $U(f)$ rotated basis
(\ref{split}) which then would lead to a logarithmic increase of the
associated entropy.

It is very gratifying that the holographic lightfront projection leads to
generalized chiral theories where such calculational techniques may be
applied. However the simplest split of longitudinal separation on one fixed
lightfront which one would use in standard chiral theories only works for
massive theories in d=1+1 when there is no chiral extension. Whenever the
horizon has a transverse extension i.e. for $d\geq1+2$ there are too many
degrees of freedom which destroy the split property\footnote{This point was
overlooked in a previous version of this work.} (even though the original
theory before the holographic projection has the split property) as it was
already evident from our previous remark that a lightcone restricted free
field is effectively a generalized free field with a continuous mass
distribution depending on $\left|  p_{\perp}\right|  $ (i.e. there exists no
density matrix description for the restricted vacuum state without the split
property). As soon as we work on two lightlike separated lightfronts and
continue the strip after crossing the edge on the second lightfront (in which
case the distance becomes spacelike instead of lightlike) we recover the split
property. This makes the application of the above Noether idea for the
construction of $U(f)$ somewhat more involved but a straightforward geometric
argument shows that the transverse factorization formula (\ref{fac}) is
recovered in the limit $\delta\rightarrow0$ when both lightfronts coalesce.
Such free field calculation combined with the fact the scale dimensions of the
generalized chiral theories on the lightfront only involve canonical
(halfinteger spin) scale dimensions strongly suggest the following picture for
the localization entropy of the strip algebras on the horizon and the nature
of the generators of the generalized chiral algebras:

\begin{itemize}
\item  The vacuum polarization strength in the limit $\delta\rightarrow0$ is
universal (logarithmic)

\item  The leading coefficient for $\delta\rightarrow0$ is proportional to the
d-2 dimensional transverse width of the strip (the area law) and the remaining
factor is ``almost'' kinematical (depends maximally on lightfront universality
classes). This suggests to split off the universally diverging factor, which
originates from the ubiquitous presence of (lightlike) vacuum polarization as
well as the area factor, which is the consequence of the degeneracy from

\item  the transverse translation symmetry. The finite numerical coefficient
which remains (which can be computed from a chiral theory on the lightray) is
the looked for (relative) area density of localization entropy. It retains the
right additive behavior for statistically independent subsystems on the
horizon, but is only determined up to a common (quantum matter independent)
factor. The remaining normalization problem requires the derivation of basic
laws of thermodynamics for this new type of horizon-associated entropy, an
issue which is outside the scope of this paper.

\item  The generating fields of the generalized chiral lightfront algebra are
of the form $\left\{  \varphi(x_{+},f)\,|\,f\in L^{2}(R^{d-2})\right\}  $
where the $\varphi$ have (half)integer scale dimensions and commute for
orthogonal $f.$
\end{itemize}

These properties, if they stand up to closer scrutiny, reinforce the quantum
mechanical transverse area structure of the above theorem. In particular the
second statement which solves the problem of converting the multiplicative
area structure of lightfront degrees of freedom into an additive area law
would be the perfect quantum match for the classical Bekenstein area law. In
fact the derivation of area proportionality for black holes entropy via
inferring thermodynamical fundamental laws on Hawking's quantum thermal
observations would be explained in terms of a totally generic local quantum
physical property which in itself does not require the presence of special
curved spacetime backgrounds.

This extremely useful role of lightfront holography in the quest for a generic
quantum Bekenstein area law is somewhat at odds with recent speculations about
the role of ``branes'' in local quantum physics. In that case the causal
shadow extensions into the ambient spacetime prevent their interpretation as
independent physical objects. This raises the question whether brane concepts
together with the closely related Kaluza-Klein reduction and many other
string-theory supported ideas can be consistent with the causal shadow
property and the control of vacuum fluctuations for small additional spatial
dimensions outside the quasiclassical approximation. For physicists who study
semiclassical aspects of branes and also believe that paradoxes and
contradictions contain the enigmatic force for progress this is an interesting situation.

It is worthwhile to point out that the use of algebraic lightfront description
goes far beyond its connection with the Bekenstein area law. It constitutes a
new tool of local quantum physics which offers all the advantages of the old
equal time canonical formalism without suffering from its short distance
limitations. Whereas the canonical formalism required the finiteness of the
wave function renormalization constants $Z$ (finiteness of the integral over
the Kallen-Lehmann spectral functions) and therefore excludes all properly
renormalizable models, the algebraic holography based on modular inclusion
avoids the ill-defined restriction of fields to space- or lightlike-
subspaces. Furthermore the chiral theories obtained by this construction are,
as already mentioned, of an extreme kinematical and universal kind. This is
because the modular inclusion method encodes the spacelike Boson/Fermi
structure in the original formulation into (half)integer dilatation spectrum
on the horizon. In fact the lightfront algebra can be described in terms of
canonical (half)integer dimension generating fields \cite{Joerss}, but they
have no obvious connection to the pointlike fields which may have generated
the original algebras; the latter have in general anomalous short distance
dimension. The scale spectrum of the kinematical chiral theory on the other
hand results from the statistics and is not related to the anomalous short
distance scale spectrum of the original theory. Although both the scaling
short distance limit and the lightfront holographic limit exhibit conformal
symmetry, the lightfront theory is much more ``universal'' than the short
distance universality classes used in the description of critical phenomena.
In addition the latter allow no mathematically controllable (since they live
in different Hilbert spaces) return to the original theory whereas the
lightfront holography reveals its massive origin upon application of the
Poincar\'{e} covariances. In short, the algebraic lightfront formalism seems
to be the long looked for dynamical ``El Dorado'': a (almost) universal
kinematical operator algebra on the lightfront (see the above conjecture about
the structure of its generating fields) which becomes reprocessed into the
rich world of quantum field theoretic models by differently acting
automorphisms. As a result of their short distance limitations neither the
canonical formalism nor the closely related Euclidean action method could play
this role, although they both served (and still serve) as useful artistic
catalyzers of thoughts about particle physics\footnote{The correctly
renormalized physical answers only fulfill Einstein causality but are neither
canonical nor Feynman-Kac representable.} (see also remarks in the last
section). But in order to convert this kinematical El Dorado into a new
powerful constructive tool of QFT one needs to get a much better conceptual
understanding and mathematical control of fuzzy acting (auto)morphisms.

What one should do is to rearrange degrees of freedom (as it already was done
in the holographic projection onto the lightfront) rather than throw away some
of them. Precisely this is achieved by the splitting method which consists in
creating a small bit finite distance between them. Here the analogy to the
inside/outside nonrelativistic quantization box (in the multiplicative
formulation of $2^{nd}$ quantization in order to make the analogy closer)
breaks down, because by removing some spatial localization region (for all
times) one is really dumping degrees of freedom. A closer examination of the
QFT situation reveals is that what splitting does is to avoid uncontrollable
divergencies which are invariably created by vacuum fluctuations if they are
forced to take place directly in a surface (or in a point in the chiral case).
Both localizations regions in this way have a fuzzy extension into the
``collar'' and this fuzzy splitting is only a rearrangement and does not
involve throwing away degrees of freedom.

\section{Comment on rotational horizons}

Let us now consider the more difficult task of a rotational symmetric double
cone and its noncompact causal disjoint. One immediately notices some
analogies to the previous planar bifurcated causal horizon. The characteristic
causal shadow of the lower lightcone horizon $h_{-}(C)$ of the unit double
cone $C$ (placed symmetric around the origin) agrees with the double cone and
its complement on the mantle of the future lightcone $h_{+}(V_{+}%
)=h(V_{+})\backslash h_{-}(C)$ is the rotational analog of the upper horizon
of the Rindler wedge. Instead of the planar longitudinal strips we now
consider strips into radial directions with a fixed space angle opening which
start at the rotational symmetric edge of bifurcation and extends to lightlike
infinity. In this way the horizon $h_{+}(V_{+})$ may be partitioned into
radial strips $\mathcal{S}_{i}$ which cast no causal shadow and whose
associated algebras mutually commute. This is the prerequisite for an angular
tensor product foliation%
\begin{align}
&  \overline{\mathcal{A}(h_{+}(V_{+}))\Omega}=H\\
&  H=\overline{\bigotimes}_{i}H_{i},\,\,H_{i}=\overline{\mathcal{A}%
(\mathcal{S}_{i})\Omega}\nonumber
\end{align}
The crucial question is whether there exists a substitute for a lightlike
translation with positive generator whose analytic properties implies the
transverse (now in the rotational sense) factorization of the vacuum into
entanglement-free strip vacua in analogy to the theorem in the previous section.

It is precisely at this point where the analogy becomes somewhat opaque. Part
of the difficulty results from the fact that the rotational causal horizon
does not come with a Killing vector field (analogous to event horizons in
curved spacetime which are not associated with Killing symmetries), except in
case the QFT is massless. In that case the Ad-action of the modular unitary
and the Tomita involution $J$ of the unit double cone corresponds to the
following point transformations (the origin is in the center of the unit
double cone \cite{Haag})%

\begin{align}
Ad\Delta^{i\frac{s}{2\pi}}  &  :x_{\pm}(s)=\frac{\left(  \cosh s\right)
x_{\pm}+\left(  \sinh s\right)  }{\left(  \sinh s\right)  x_{\pm}+\left(
\cosh s\right)  },\,\,x_{\pm}\equiv x^{0}\pm\left|  \vec{x}\right|
\label{mod}\\
J  &  :x^{i}\rightarrow-\frac{x^{i}}{x^{2}},\,\,x^{0}\rightarrow\frac{x^{0}%
}{x^{2}},\,\ \,-1<x_{\pm}<1\nonumber
\end{align}
Although an explicit description of the corresponding fuzzy modular
transformations in the massive case is presently not possible\footnote{The
conjecture that these transformation correspond to double cone (and its causal
disjoint) support preserving test function transformations of the
pseudo-differential kind \cite{S-W} still stands.}, the modular objects become
geometric in the holographic projection in which the double cone is projected
onto its lower lightcone mantle and the spacelike disjoint projects on the
infinite complementary outside part of the mantle.

\begin{remark}
The restriction of the double-cone localized free massive field to the lower
horizon (i.e. to the mantle of that part of the forward lightcone whose causal
shadow is the double cone) follows the method used for restricting free field
operators to $LF_{+}$ \cite{Hor}$:$ the limit $x_{-}=x^{0}-\left|  \vec
{x}\right|  \rightarrow0$ in the plane wave factor (with the origin at the
lower apex of the double cone) is compensated by an $\ln x_{-}$-increase in
the spatial rapidity $\chi;$ the creation/annihilation operators remain
unaffected. The resulting free field restriction is that of a massless field
(the mass only remains as a scale factor in the exponential), however the
physical mass is re-activated by acting with Poincar\'{e} transformations on
the lightcone restriction.
\end{remark}

It is believed that the holographic lightcone projection of all massive
theories admits this geometric modular group action%
\begin{align}
Ad\Delta^{i\frac{s}{2\pi}} &  :x_{+}(s)=\frac{\left(  \cosh s\right)
x_{+}+\left(  \sinh s\right)  }{\left(  \sinh s\right)  x_{+}+\left(  \cosh
s\right)  }\,\\
x_{-} &  =0,\,\,\,x_{+}\equiv x^{0}+\left|  \vec{x}\right|  ,\,\,\left|
x_{+}\right|  \leq1
\end{align}
and on the holographic projection of its causal disjoint $x_{-}=0,\,\,x_{+}%
\geq1.$ This leads to the interesting conjecture that the theorem about
modular inclusion of the previous section may have a partially geometric
counterpart in which two fuzzy modular groups\footnote{It has been conjectured
that the action of \ fuzzy modular actions on testfunctions should be
described by a special type of pseudo-differential operators which leaves
their given $\mathcal{O}$-support and its causal disjoint invariant
\cite{S-W1} but unforunately no progress has been obtained on this point.} may
nevertheless lead to a geometric Borchers pair (positive lightlike
translation, dilation) on the lightcone horizon. Further work is necessary to
clarify the situations of causal (event) horizons without Killing symmetries.

\section{Concluding remarks}

My use of the adjective ``paradigmatic'' in the title requires more justification.

In most uses of methods of AQFT there was always a way to see things (perhaps
not in the most elegant fashion) in terms of field-coordinatizations. In the
case at hand, as a result of severe short distance limitations which (except
for free fields) rule out to define holographic lightfront projections by
restricting pointlike fields, this is not possible.

The main point in the present work was to establish a characteristic property
of a new algebraic approach to ``holographic lightfront projection''
(including a precise definition), namely its perfect transverse quantum
mechanical aspect. According the best of my knowledge this is the only known
case where Galilei-invariant vacuum-polarization-free quantum mechanics
appears within QFT as a rigorous structural property i.e. without any
nonrelativistic approximation $v<<c$.

The prerequisite for this extraordinary phenomenon is the spacetime
arrangement suffered by local quantum physical degrees of freedom on
non-globally hyperbolic manifolds. Situations as the AdS-CQFT correspondence
\cite{Mald} (also often referred to as AdS-CQFT holography) (in which the
local form of causal propagation is still maintained\footnote{In a recent
paper \cite{Re2} it was shown that the string-theory induced duality relation
of AdS with CQFT is just an alternative formal reformulation of the field
theoretic restriction at the conformal boundary at infinity (which permits a
rigorous algebraic presentation \cite{Re1}).}), are not radical enough for
obtaining such a situation. One really needs the absence of causal shadows as
encountered on causal (event) horizons where only \textit{seminifinite
regions} which extend to lightlike infinity are capable of casting causal
shadows into the ambient spacetime. The operator algebras associated with such
seminifinite regions are the only ones which geometrically communicate between
the lightfront and its ambient spacetime.

This use of change of spacetime indexing as a tool of structural investigation
in QFT links up very nicely with a recent stunning discovery about the true
nature of QFT \cite{Br}. Whereas prior to that observation the curved
spacetime aspect were always considered as part of the specification of the
model, the new view of QFT is that of a functor between the categories of
globally hyperbolic Lorentz signature (sub)manifolds (with the arrows being
isometric embedding) and that of operator algebras (with the arrows being
morphisms). In other words the curved spacetime aspect is not an additional
property imposed on a QFT but belongs to the very definition of what
constitutes a QFT. In order to illustrate this paradigmatic change of thinking
about QFT let me be somewhat metaphoric in my analogies and compare the
abstract form of QFT with a kind of unstructured mold of degrees of freedom
(being structured in different ways by different space-time
indexing\footnote{An illustration would be the abstract Weyl algebra as an
operator-valued functor on abstract Hilbert spaces versus the concrete
labeling by identifying the abstract space as a concrete function space and
using the localized subspaces of the latter to generate a concrete net
structure on the former.})\ similar to stem-cells and their versatile use for
the generation of different biological tissue. The spacetime acting as agents
for structuring the degree of freedoms by forcing them to align in a certain
way. If anything, the present modular supported holography deepens this
surprising new view in that it permits to change the spacetime indexing in an
even more (not generally understood) radical way by allowing lower dimensional
sets. It would be nice if algebraic holographic projections onto horizons
could be viewed as a generalization of spacetime reprocessing to situations in
which there are no connecting local diffeomorphisms as in \cite{Br}.

Extrapolating from properties of interaction-free lightfront holography in the
present work, one expects the existence of a relative area density of
localization entropy associated with the horizon. There are two prerequisites
which would insure the validity of an area law as the local quantum
counterpart of the classical Bekenstein law for black hole entropy. The
``split entropy'' of the vacuum restricted to the algebra of a split interval
in a chiral theory associated to the lightfront should increase with the
splitting distance $\varepsilon\rightarrow0$ in a universal way so that the
ratio between two models with different matter content but the same geometric
split situation stays finite. For the normalization of such a relative entropy
one needs a second prerequisite which is the validity of a fundamental
thermodynamic law which relates the entropy to other already normalized quantities.

In contrast to this envisaged generic area behavior, there have been
derivations of area laws for some kind of horizon-affiliated entropy (whose
precise local quantum physical interpretation is not clear) in models of
string theory \cite{String} or by imposing a classical Virasoro structure on
horizons and quantizing \cite{Carlip}, or last not least by rewriting
canonical classical GR variables into loop-like variables \cite{Loop}.
Although the conditions of their derivation seems to be too special in order
to account for the apparently universal Bekenstein area law (unless one
believes in the TOE doctrine for those models), they are not necessarily
contradicting the present lightfront holography mechanism. It is conceivable
that a local quantum phenomenon, which is generic in the algebraic setting but
at the same time not easily accessible in the more geometric standard
formulation of QFT, may well present itself in special geometric situations in
a more manifest fashion.

The present view de-emphasizes somewhat the importance of black holes as a
direct entrance ticket into QG. But not completely, since there is the very
interesting message of an apparent very \textit{deep connection of thermal
physics with geometry} at the place where one wants to see new quantum gravity
degrees of freedom. This has already been foreshadowed by recent results on
the construction of external and internal symmetries and spacetime geometry
from the relative position of operator algebras \cite{Wies}\cite{Kaehler}%
\cite{Bu}, and in particular the emergence of infinite dimensional fuzzy
analogs of diffeomorphism groups (including the Poincar\'{e} and conformal
diffeomorphisms) from modular inclusions and intersections of algebras point
into the same direction \cite{S1}.

\textbf{Acknowledgment}: I am thankful to Detlev Buchholz for directing my
attention to the almost forgotten early work of W. Driessler. I also
acknowledge critical remarks of K.-H. Rehren which led to some modifications
of the first version.

\section{Appendix: Lightfront restriction}

We start with a d=1+1 massive free field in the momentum space rapidity
description
\begin{align}
A(x)  &  =\frac{1}{\sqrt{2\pi}}\int\left(  e^{-ipx}a(\theta)+e^{ipx}a^{\ast
}(\theta)\right)  d\theta\\
p  &  =m(ch\theta,sh\theta)\nonumber
\end{align}
The positive part $x_{+}\equiv t+x>0$ of the light ray $x_{-}=t-x=0$ which is
the upper horizon of the wedge $t^{2}-x^{2}<0,$ $x>0$ is approached by taking
the $r\rightarrow0,\,\chi=\hat{\chi}-ln\frac{r}{r_{0}}\rightarrow\hat{\chi}$
$+\infty$ in the x-space rapidity parametrization
\begin{align}
&  x=r(sh\chi,ch\chi),\,\,x\rightarrow(x_{-}=0,x_{+}\geq0,\text{ }finite)\\
&  A(x_{+},x_{-}\rightarrow0)\equiv A_{LF}(x_{+})=\frac{1}{\sqrt{2\pi}}%
\int\left(  e^{-ip_{-}x_{+}}a(\theta)+e^{ip_{-}x_{+}}a^{\ast}(\theta)\right)
d\theta\nonumber\\
&  =\frac{1}{\sqrt{2\pi}}\int\left(  e^{-ip_{-}x_{+}}a(p)+e^{ip_{-}x_{+}%
}a^{\ast}(p)\right)  \frac{dp}{\left|  p\right|  }\nonumber
\end{align}
where the last formula serves to make manifest that the limiting $A_{LF}%
(x_{+})$ field is a chiral conformal (gapless $P_{-}$ spectrum) field; the
mass $m$ in the exponent $p_{-}x_{+}=mr_{0}e^{\theta}e^{-\hat{\chi}}$ is a
dimension preserving parameter which has lost its physical significance of a
mass gap; the physical mass is related to the gap in the $P_{-}\cdot P_{+}$
spectrum of the ambient d=1+1 theory. Note that unlike the scaling limit, the
lightfront only effects the numerical factors and not the Fock space operators
$a^{^{\#}}(\theta).$

In this formal argument using pointlike fields one has to keep in mind that a
free massless $dimA=0\,$scalar field is only defined on a restricted space of
localized test functions. This restriction arises automatically if one
performs the limit in a more careful way including smearing with
wedge-supported test functions. With $supp\tilde{f}\in W,\,\tilde{f}$ real,
one obtains
\begin{align}
&  \int A(x_{+},x_{-})\tilde{f}(x)d^{2}x=\int_{C}a(\theta)f(\theta)=\int
A_{LF}(x_{+})\tilde{g}(x_{+})dx_{+},\,\tilde{g}\text{ }real\\
&  \tilde{f}(x)=\int_{C}e^{ip(\theta)x}f(\theta)d\theta,\,\,\tilde{g}%
(x_{+})=\int e^{ipx_{+}}g(p)\frac{dp}{\left|  p\right|  }=\int_{C}%
f(\theta)e^{ip_{-}(\theta)x_{+}}d\theta\nonumber
\end{align}
These formulas, in which a contour $C$ appears, require some explanation. The
on-shell character of free fields restricts the Fourier transformed test
function $f(p)$ to their mass shell values with the backward mass shell
corresponding to the rapidity on the real line shifted downward by $-i\pi$
\[
f(p)|_{p^{2}=m^{2}}=\left\{
\begin{array}
[c]{c}%
f(\theta),\,\,p_{0}>0\\
f(\theta-i\pi),.\,\,\,p_{0}<0
\end{array}
\right.
\]
and the wedge support property is equivalent to the analyticity of $f(z)$ in
the strip -$i\pi<\operatorname{Im}z<0$. The integration path $C$ consists of
the upper and lower rim of this strip and hence corresponds to the
negative/positive frequency part of the Fourier transform. The vanishing of
the massless test function $g(p)$ is nothing but the statement that the wave
function $f(\theta)$ considered as a zero mass shell restriction of a massless
test function $g(p)$ must have the vanishing property at $p=0$ since this
corresponds to the value $\theta=-\infty$ of the square integrable
$f(\theta)$
\begin{align}
f(p)|_{p^{2}=m^{2},p_{0}>0}  &  \frac{dp}{\sqrt{p^{2}+m^{2}}}=f(\theta
)d\theta\equiv g(\theta)d\theta=g(p)|_{p^{2}=0,p_{0}>0}\frac{dp}{\left|
p\right|  }\\
&  \curvearrowright g(p=0)=0,\,or\,\int\tilde{g}(x_{+})dx_{+}=0\nonumber
\end{align}
with a similar formula for negative $p_{0}$ and the corresponding $\theta
$-values at the lower rim. Note that this argument excludes the ``naive''
(unrestricted) $dimA=0$ zero mass free field which would violate Huygens
principle, require indefinite metric and violate cluster properties (and hence
contradict the main theorem in section 3).

The derivation of these formulas for $d>1+1$ becomes word for word the same if
one introduces the effective mass $m_{eff}=\sqrt{m^{2}+p_{\perp}^{2}\text{ }}$
where $p_{\perp}$ is transverse to the lightray momentum. In that case one
should do the calculation on the basis of longitudinal/transverse product test
functions $f(p)=f_{l}(p_{l})f_{t}(p_{\perp})$ and extend linearly after the
lightfront limit has been performed. The $f_{l}$ passes to a massless $g_{l}$
just as in the d=1+1 case. In terms of the two-point function the result is%

\begin{align}
&  \left\langle A_{LF}(f_{l}f_{\perp})A_{LF}(f_{l}^{\prime}f_{\perp}^{\prime
})\right\rangle =\int\bar{f}_{l}(p_{l})f^{\prime}(p_{l})\frac{dp_{l}}{2\left|
p_{l}\right|  }\int\bar{f}_{\perp}(p_{\perp})f_{\perp}^{\prime}(p_{\perp
})d^{2}p_{\perp}\nonumber\\
&  \left[  A_{LF}(x_{+},x_{\perp}),A_{LF}(x_{+}^{\prime},x_{\perp}^{\prime
})\right]  =i\Delta(x_{+}-x_{+}^{\prime})_{m=0}^{{}}\delta(x_{\perp}-x_{\perp
}^{\prime})
\end{align}
i.e. the commutation of the transverse part is like that of \ Schr\"{o}dinger
field. In fact the analogy to QM is much stronger since the vacuum does not
carry any transverse correlation at all. The total correlationless property is
best seen in the algebraic transcription using the Weyl generators
\begin{align}
\left\langle W(f_{l},f_{\perp})W(f_{l}^{\prime},f_{\perp}^{\prime
})\right\rangle  &  =\left\langle W(f_{l},f_{\perp})\right\rangle \left\langle
W(f_{l}^{\prime},f_{\perp}^{\prime})\right\rangle \text{ }if\,\,suppf_{\perp
}\cap suppf_{\perp}^{\prime}=\emptyset\label{Weyl}\\
W(f_{l},f_{\perp})  &  =e^{iA_{LF}(f_{l}f_{\perp})}\nonumber
\end{align}
i.e. the vacuum behaves quantum mechanically in the transverse direction.

In the interacting case when the integral over the Kallen-Lehmann spectral
function diverges $\int\rho(\kappa)d\kappa=\infty$ one has to use the
algebraic modular inclusion method instead of the lightfront restriction formalism.

The quantum mechanical transvers aspect of LF-QFT is corroborated by the fact
that its 7-parametric symmetry group (which is a subgroup of the 10-parametric
Poincar\'{e} group) contains a copy of the transvers Galilei group
\cite{S2}\cite{Hol}.


\begin{thebibliography}{99}
\bibitem{LKS}H. Leutwyler, J. R. Klauder and L. Streit, Nuovo Cim.
\textbf{66A}, (1970) 536

\bibitem{Ho}G. 't Hooft, in Salam-Festschrift, A. Ali et al. eds., World
Scientific 1993, 284

\bibitem{Suss}L. Susskind, J. Math. Phys. \textbf{36}, (1995) 6377

\bibitem{Wald}R. M. Wald, \textit{Quantum Field Theory in Curved Spacetime and
Black Hole Thermodynamics}, The University of Chicago Press 1994, and
references cited therein

\bibitem{S2}B. Schroer, \textit{Lightfront holography and area density of
entropy associated with localization on wedge-horizons}, and references
therein, hep-th/0208113 to appear in IJMPA

\bibitem{Re}K.-H. Rehren, Ann. Henri Poincar\'{e} \textbf{1}, (2000) 607

\bibitem{Bu}D. Buchholz, \textit{Algebraic Quantum Field Theory: A Status
Report}, hep-th/00112701

\bibitem{Hol}S. Hollands and R. Wald, Commun. Math. Phys. \textbf{223}, (2001) 289

\bibitem{Br}R. Brunetti, K. Fredenhagen and R. Verch, \textit{The Generally
Covariant Locality Principle}-\textit{A New Paradigm for Local Quantum field
Theory}, hep-th/01122000 and references on prior work

\bibitem{Haag}R. Haag, \textit{Local Quantum Physics}, Springer Verlag (1992)

\bibitem{S-matrix}B. Schroer, \textit{Uniqueness of Inverse Scattering Problem
in Local Quantum Physics}, hep-th/0106066

\bibitem{Bo-JMP}H. J. Borchers, J. Math. Phys. \textbf{41}, (2000) 3604

\bibitem{Hal}H. Halvorson, \textit{Reeh-Schlieder Defeats Newton-Wigner: On
alternative localization schemes in relativistic quantum field theories} and
references therein, quant-ph/0007060

\bibitem{qu}B. Schroer, \textit{Basic Quantum Theory and Measurement from the
Viewpoint of Local Quantum Physics}, in Trends in Quantum Mechanics, page 274,
ed. H.-D.Doebner, S. T. Ali , M. Keyl and R. F. Werner, World Scientific 2000, quant-ph/9904072

\bibitem{St-Wi}R. F. Streater and A. S. Wightman, PCT, Spin\&Statistics and
all That, Benjamin 1964

\bibitem{H-S}R. Haag and B. Schroer, J. Math. Phys. \textbf{3}, (1962) 248

\bibitem{Wies}H.-W. Wiesbrock, Comm. Math. Phys. \textbf{158}, (1993) 537

\bibitem{Guido}D. Guido, R. Longo and H.-W. Wiesbrock, Commun. Math. Phys.
\textbf{192}, (1998) \ 217

\bibitem{Hei}Selected papers on \textit{Quantum Electrodynamics}, edited by J.
Schwinger, Dover Publications Inc. New York 1958

\bibitem{Driessler}W. Driessler, Acta Physica Austriaca \textbf{46}, (1977)
163, and references therein

\bibitem{GLRV}D. Guido, R. Longo, J.E. Roberts and R. Verch, Charged sectors,
spin and statistics in quantum field theory on curved spacetimes, \ math-ph/9906019

\bibitem{BBS}H.J. Borchers, D. Buchholz and B. Schroer, Commun. Math. Phys.
\textbf{219}, (2001) 125, and prior references quoted therein.

\bibitem{JPL}B. Schroer. Phys. Lett. \textbf{B 494} (2000) 124

\bibitem{Schroer}B. Schroer, J. Math. Phys. \textbf{41}, (2000) 3801 and
ealier papers of the author quoted therein

\bibitem{S-W}B. Schroer and H.-W. Wiesbrock Rev. Math. Phys. \textbf{12},
(2000) 461

\bibitem{Wigner}E. P. Wigner, Ann. Math. \textbf{40}, (1939) 149

\bibitem{Borchers}H. J. Borchers, Commun. Math. Phys. \textbf{2}, (1966) 49

\bibitem{G-L}R. Brunetti, D. Guido and R. Longo, Commun. Math. Phys.
\textbf{156}, (1993) 201

\bibitem{S}B. Schroer, Phys. Lett. \textbf{B} \textbf{506}, (2001) 337

\bibitem{D-L}S. Doplicher and R. Longo, Invent. math. \textbf{75}, (1984) 493

\bibitem{Narn}H. Narnhofer, in \textit{The State of Matter}, ed. by M.
Aizenman and H. Araki (Wold-Scientific, Singapore) 1994

\bibitem{ADFL}C. D'Antoni, S. Doplicher, K. Fredenhagen and R. Longo, Commun.
Math. Phys.\textbf{110}, (1987) 325

\bibitem{Joerss}M. J\"{o}rss, Lett. Math. Phys. \textbf{38}, (1996) 252

\bibitem{Hor}B. Schroer, \textit{Lightfront Formalism versus
Holography\&Chiral Scanning}, hep-th/0108203

\bibitem{Mald}J. M. Maldacena, Adv. Theor. Math. Phys.2, (1998) 231

\bibitem{Re1}K.-H. Rehren, Phys. Lett.B \textbf{493}, (2000) 383

\bibitem{Re2}M. Duetsch and K.-H. Rehren, \textit{A comment on the dual field
in the AdS-CFT correspondence}, hep-th/0204123, to appear in Lett. Math. Phys.

\bibitem{String}G. Horowitz, \textit{Quantum Gravity at the Turn of the
Millenium}, hep-th/0011089

\bibitem{S-W1}B. Schroer and H-W Wiesbrock, Rev. Math. Phys. \textbf{12},
(2000) 139

\bibitem{Carlip}S. Carlip, Nucl.Phys.Proc. Suppl.\textbf{88}, (2000) 10, gr-qc/9912118

\bibitem{Loop}T. Thiemann, \textit{An Introduction to Modern Canonical Quantum
General Relativity}, gr-qc/0110034

\bibitem{Kaehler}R. Kaehler and H.-W. Wiesbrock, JMP \textbf{42}, (2000) 74

\bibitem{S1}L. Fassarella and B. Schroer, Phys. Lett. B \textbf{538}, (2002) 415
\end{thebibliography}
\end{document}